%% file: eprint.tex
\documentclass[12pt]{article}
\usepackage{graphicx}
\usepackage[labelformat=simple]{subcaption}
\usepackage[utf8]{inputenc}
\usepackage{amsmath}

\newcommand\pubnumber{}
\newcommand\pubdate{\today}

\def\institute{LIP\\
Universidade do Minho, Braga, Portugal}
\def\support{\footnote{Work supported by FCT through grant PD/BD/135435/2017, through projects CERN/FIS-PAR/0008/2017 and PTDC/FIS-PAR/29147/2017, COMPETE2020-Portugal2020, POSI through project POCI-01-0145-FEDER-029147.\\Copyright 2018 CERN for the benefit of the ATLAS Collaboration. Reproduction of this article or parts of it is allowed as specified in the CC-BY-4.0 license.}}

\def\Title#1{\begin{center} {\Large #1 } \end{center}}
\def\Author#1{\begin{center}{ \sc #1} \end{center}}
\def\Address#1{\begin{center}{ \it #1} \end{center}}

\newcommand\pubblock{\rightline{\begin{tabular}{l} \pubnumber\\
         \pubdate  \end{tabular}}}
\newenvironment{Abstract}{\begin{quotation}  }{\end{quotation}}
\newenvironment{Presented}{\begin{quotation} \begin{center} 
             PRESENTED AT\end{center}\bigskip 
      \begin{center}\begin{large}}{\end{large}\end{center} \end{quotation}}

\input econfmacros.tex

\newcommand{\pt}{\ensuremath{p_{\text{T}}}}
\newcommand*{\Fig}[1]{Figure~#1}

\newcommand{\HT}{\ensuremath{H_{\text{T}}}}
\newcommand{\ST}{\ensuremath{S_{\text{T}}}}
\newcommand{\ttbar}{t\bar{t}}	
\newcommand{\bjet}{$b$-tagged jet }
\newcommand{\dilres}{2$\ell$ 0-1J}
\newcommand{\dilboos}{2$\ell$ $\geq2$J}
\newcommand{\trilep}{3$\ell$}
\begin{document}
\begin{titlepage}
\pubblock

\vfill
\Title{Search for pair-production of vector-like quarks in final states with at least one $Z$ boson decaying into a pair of electrons or muons in $pp$ collision data collected with the ATLAS detector at $\sqrt{s}$ = 13 TeV}
\vfill
\Author{Tiago Vale\support,\\on behalf of the ATLAS Collaboration}
\Address{\institute}
\vfill
\begin{Abstract}
A search for pair-produced vector-like quarks decaying into a $Z$ boson and a third generation quark using 2015 and 2016 $pp$ collision data at $\sqrt{s}~=~13$~TeV collected by the ATLAS experiment at the Large Hadron Collider, corresponding to an integrated luminosity of 36.1 fb$^{-1}$, is presented. No significant excess over the Standard Model expectation was found so lower exclusion limits at 95\% confidence level were set on the vector-like quark masses at 1030 (1210) GeV for singlet (doublet) $T$ and 1010 (1140) GeV for singlet (doublet) $B$.
\end{Abstract}
\vfill
\begin{Presented}
$11^\mathrm{th}$ International Workshop on Top Quark Physics\\
Bad Neuenahr, Germany, September 16--21, 2018
\end{Presented}
\vfill
\end{titlepage}
\def\thefootnote{\fnsymbol{footnote}}
\setcounter{footnote}{0}

\section{Introduction}
%The Standard Model (SM) of particle physics is a often regarded as a low-energy approximation of a more fundamental theory, as several questions remained unanswered by it. One of these is the hierarchy problem, which arises when the SM is extrapolated to higher energies and fine-tuning is necessary to correct for the Higgs boson self-energy divergence. CITE 1 from VLQ paper. Many models that try to fix this problem do so by trying to reduce the top quark contribution to this divergence, given that it is the largest of all SM particles. This is generally done with the addition of a top partner. Models in which this top partner is fermionic often introduce vector-like quarks (VLQ) \cite{aguilar}. Examples of these models include Little Higggs and Composite Higgs CITE where the Higgs boson is a pseudo Nambu-Goldstone boson from a new broken global symmetry.\par 
Vector-like quarks (VLQ) are color-triplet spin-1/2 fermions whose left- and right-hand component transform in the same way under SU(2)$_L \times$ U(1)$_Y$ (from which they are given the name vector-like). The electric charge can be the SM-like with +2/3$e$ ($T$) and ${-1/3}e$ ($B$) or exotic with +5/3$e$ (X) and ${-4/3}e$ (Y), where $e$ is the elementary charge. This search \cite{analysis} only takes into account the SM-like electrically charged quarks. Given the assumption that these quarks can couple to the Higgs boson a limitation in their possible isospin quantum numbers is set. They can be singlets ($T$ or $B$), doublets ($X T$, $T B$ or $B Y$) or triplets ($X T B$ or $T B Y$). They are predicted in many models that introduce fermionic top partners \cite{aguilar}.\par
The present proceedings focus on events consistent with pair-produced VLQ with decays through a $Z$ boson and a third generation quark, using data collected by the ATLAS experiment\cite{atlas} in 2015 and 2016 corresponding to 36.1 fb$^{-1}$.
\section{Analysis strategy}
\begin{figure}[htb]
\centering
\includegraphics[height=1.5in]{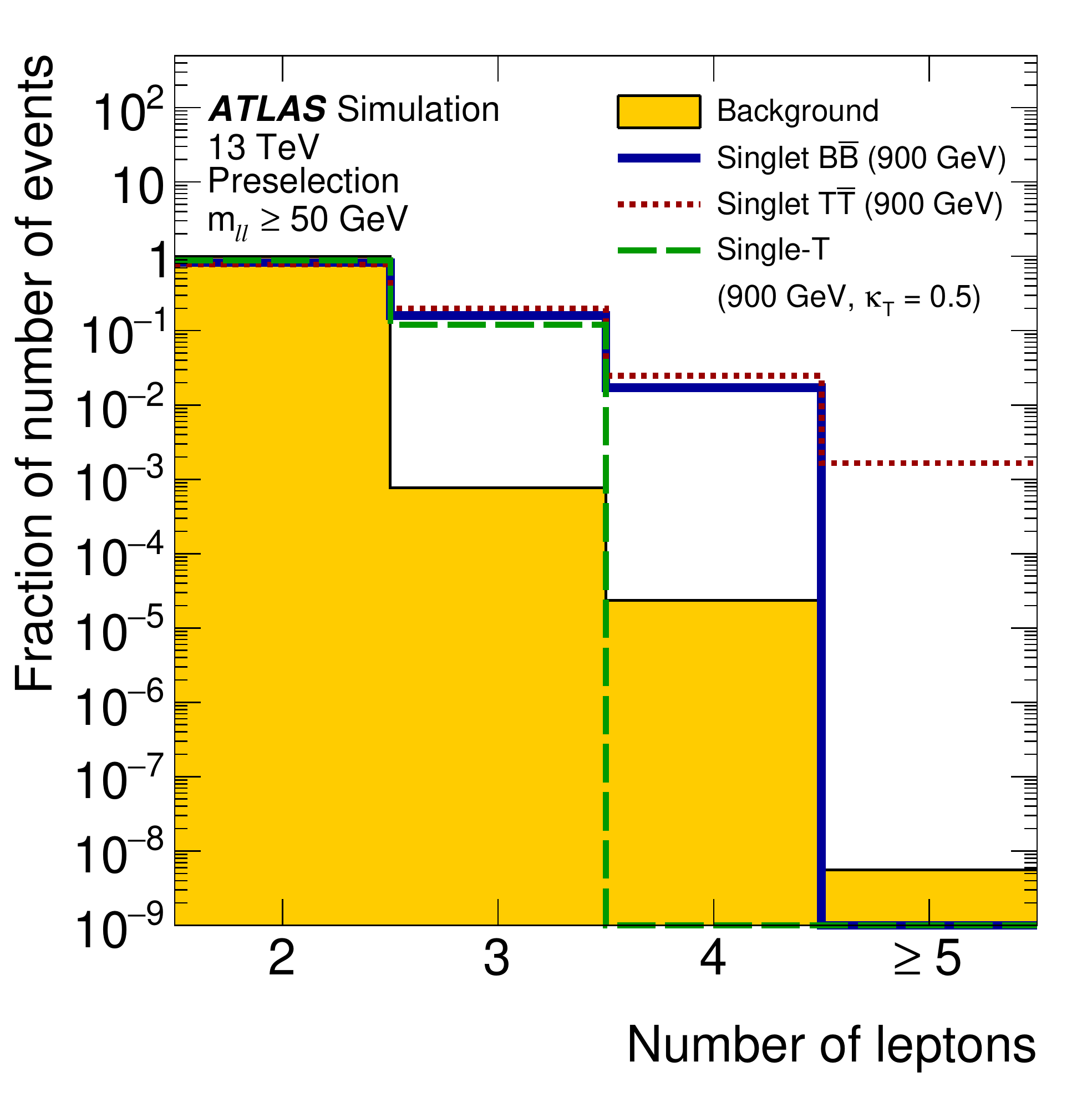}
\includegraphics[height=1.5in]{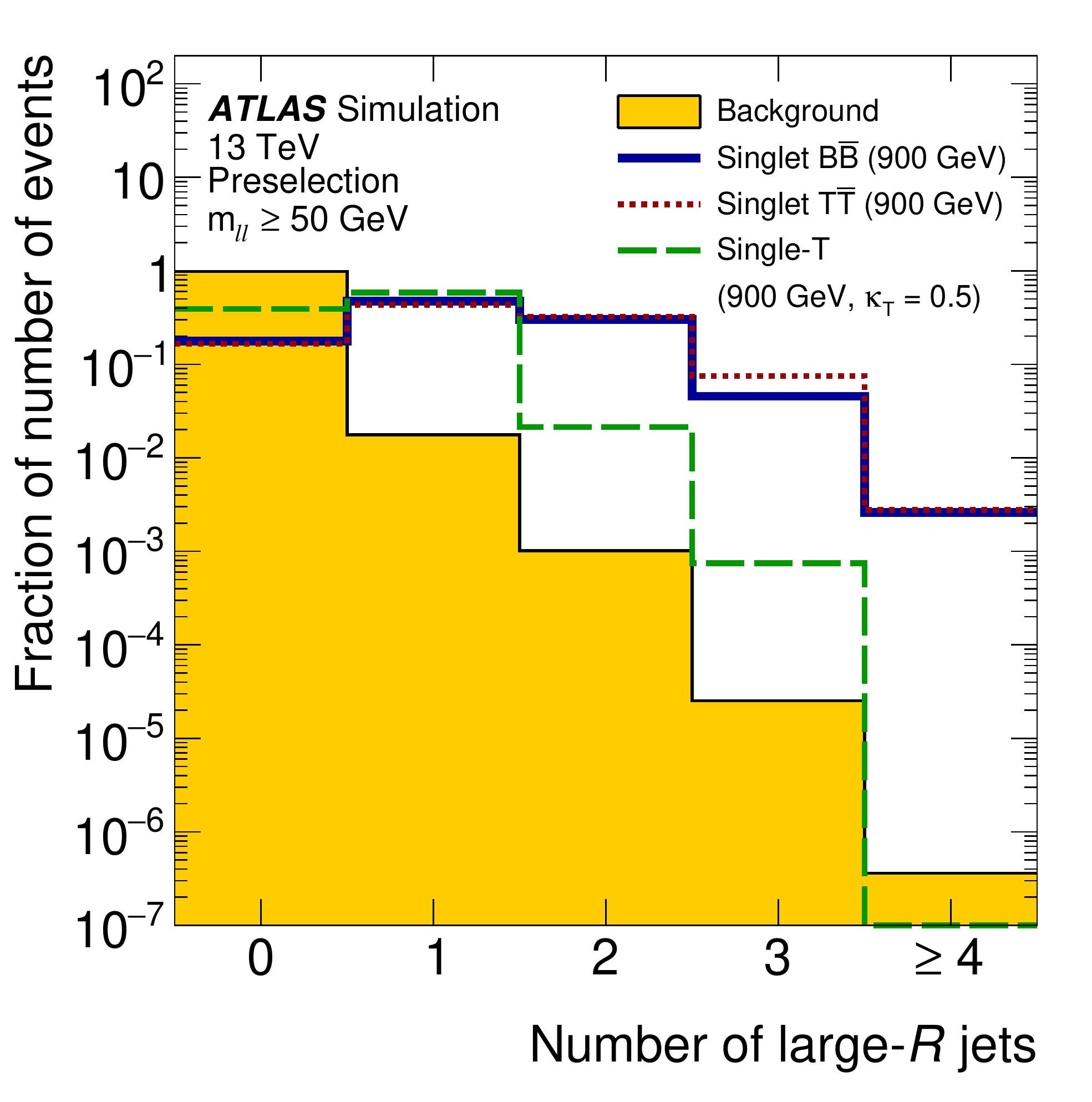}
\caption{Distributions of the number of leptons (left) and the number of large-R jets (right). All background processes are added and all distributions are normalized to unit area. The last bin contains overflow.}
\label{fig:shape_plots}
\end{figure}

VLQ pair-production is expected to produce high multiplicity final states when compared to background processes. In Figure \ref{fig:shape_plots}, unit normalized plots with the multiplicity of leptons on the left and large-R jets on the right are shown. In this figure the shape differences between signal and background are noticeable. Focusing on the lepton multiplicity it can be seen that the signal-to-background ratios in the first bin and the further ones are significantly different, thus a split is made: the dilepton channel (exactly two leptons) will benefit from an increase in statistics whereas the trilepton channel (at least three leptons) will have a much better signal-to-background ratio.\par 
When the large-R jet multiplicity is considered a similar conclusion is reached. Therefore another split is made: the dilepton channel will focus separately in events with 0 or 1 large-R jet (with the label \dilres) and events with at least 2 large-R jets (\dilboos). The trilepton channel (\trilep) will make no cuts in this category. 
%The resulting splits are summarized in Figure \ref{fig:diagrams} where a sketch of the processes can be seen.
%\begin{figure}[h]
%\centering
%\includegraphics[width=0.3\textwidth]{figures/fig_01a.pdf}
%\includegraphics[width=0.3\textwidth]{figures/fig_01b.pdf}
%\includegraphics[width=0.3\textwidth]{figures/fig_01c.pdf}
%\caption{Sketches of the processes searched in this analysis.}
%\label{fig:diagrams}
%\end{figure}
\section{Signal regions}
%Having split the pair-production search we optimize each channel. The goal is to have three separately optimized channels, studying the different background compositions and reaching the optimal selection in each, which will provide a better sensitivity than doing an inclusive pair-production search.\par 
Events corresponding to VLQ with high masses are expected to be boosted with respect to background processes, which motivates a selection based on these differences. Therefore targeting the transverse momentum ($\pt$) of particles should allow to extrapolate these differences.\par
\begin{figure}[h]
\centering
\begin{subfigure}[b]{0.51\textwidth}
\includegraphics[width=0.491\textwidth]{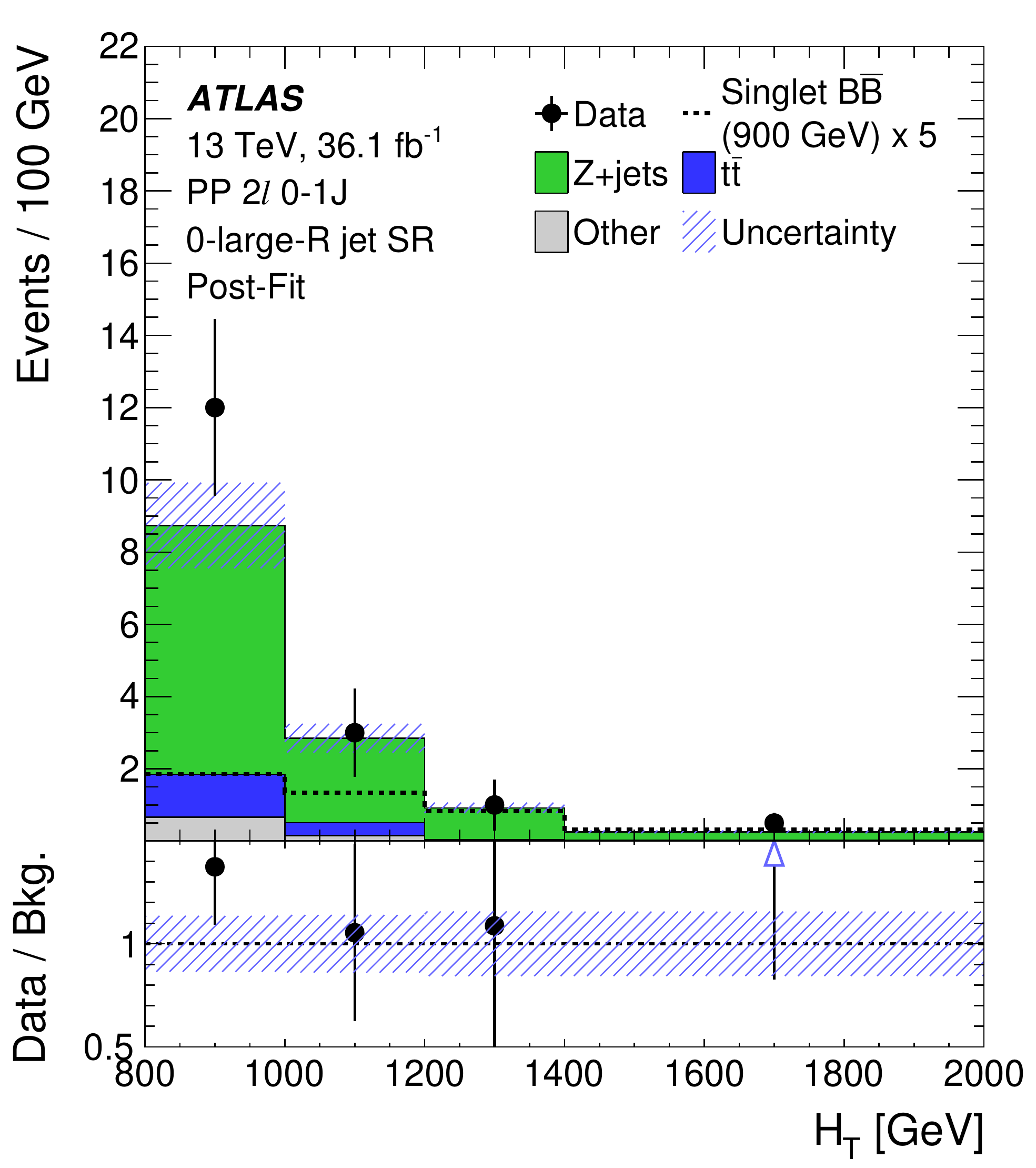}
\includegraphics[width=0.491\textwidth]{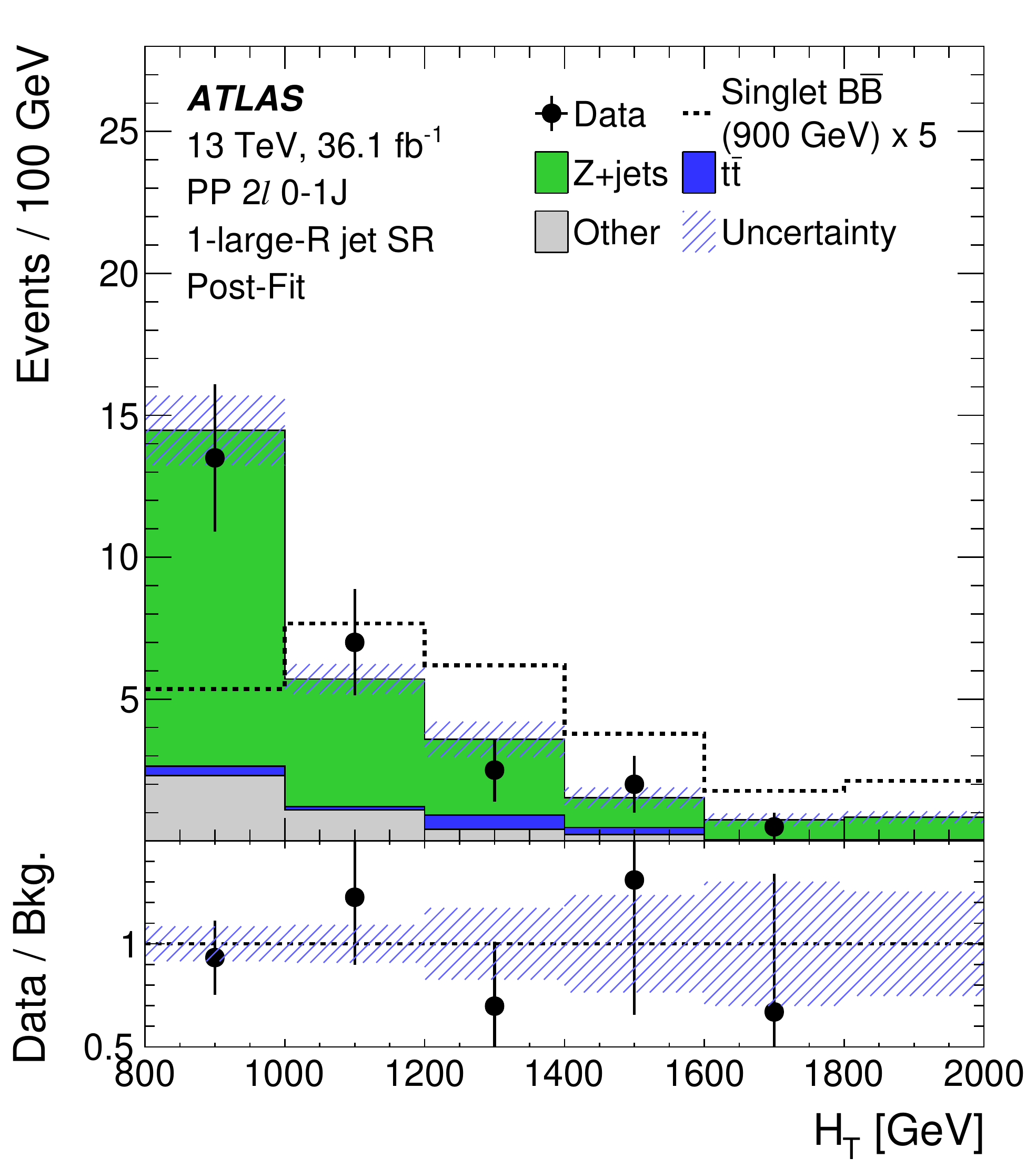}
\caption{}
\label{fig:dilres}
\end{subfigure}\\
\begin{subfigure}[b]{0.25\textwidth}
\includegraphics[width=\textwidth]{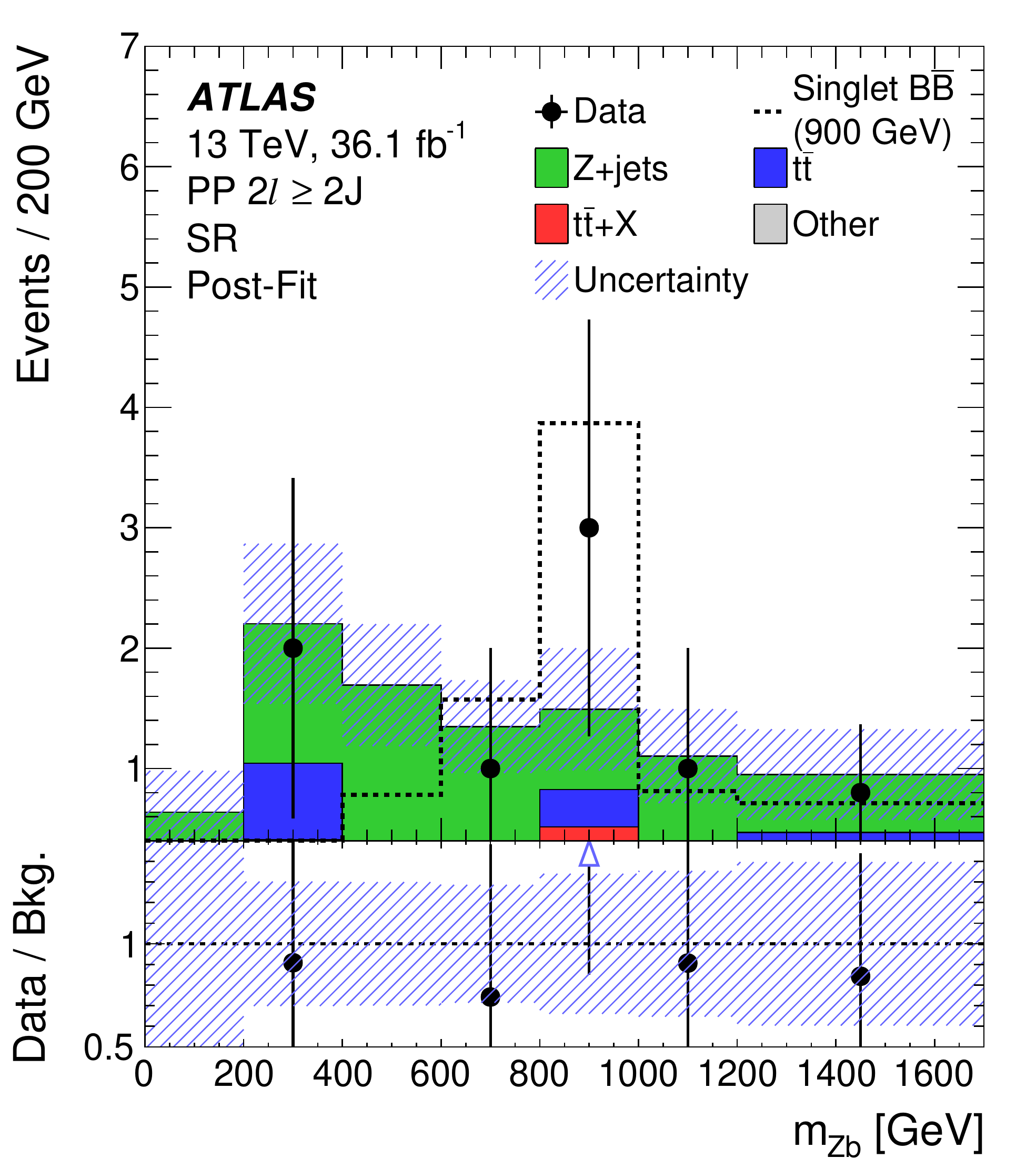}
\caption{}
\label{fig:dilboos}
\end{subfigure}
\begin{subfigure}[b]{0.25\textwidth}
\includegraphics[width=\textwidth]{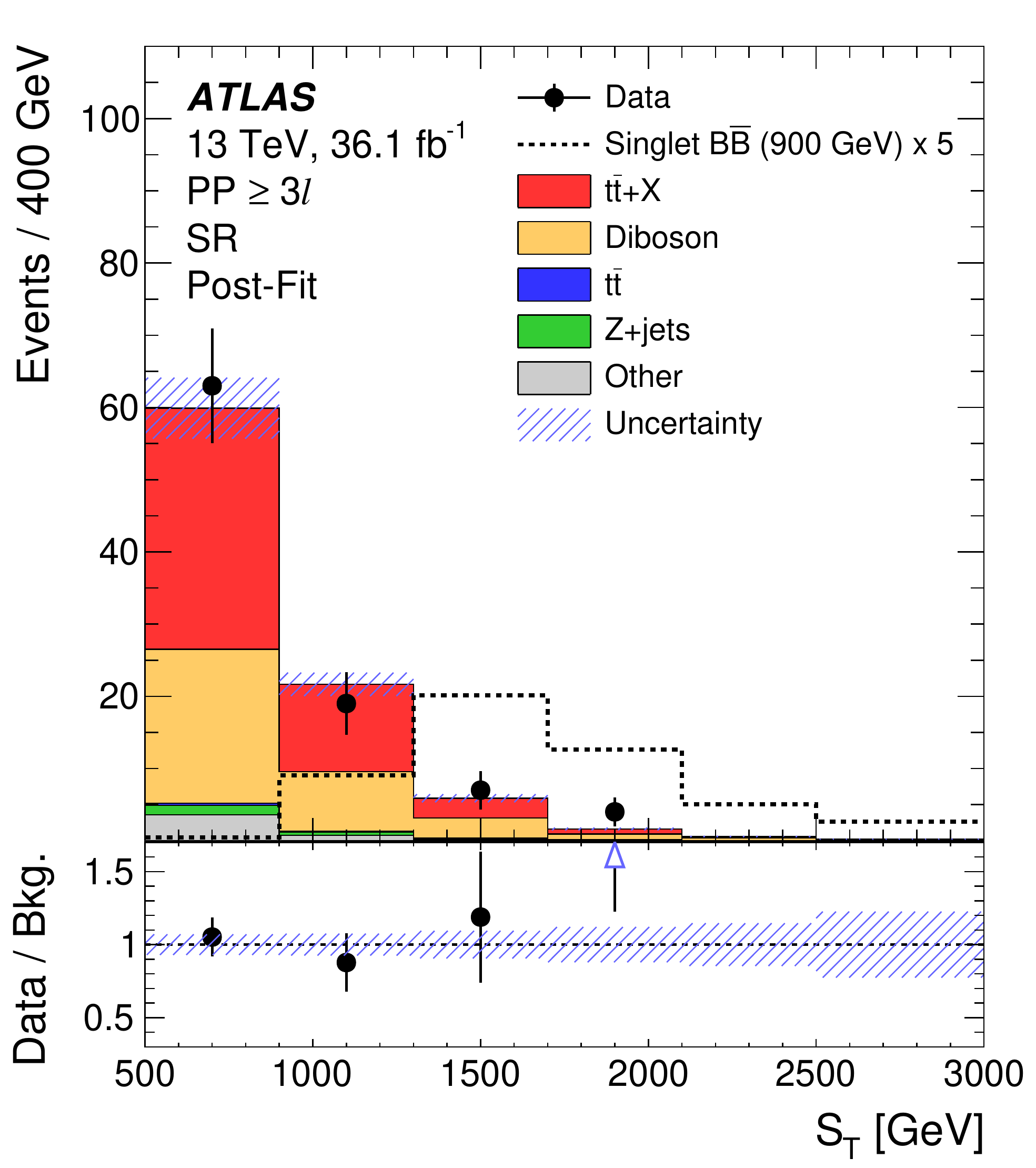}
\caption{}
\label{fig:tril}
\end{subfigure}
\caption{\subref{fig:dilres} Scalar sum of jets transverse momenta in the 2$\ell 0-1$J channel's signal regions. \subref{fig:dilboos} Invariant mass of the $Z$ boson candidate and the jet with the highest traverse momentum in the 2$\ell \geq2$J channel's signal region. \subref{fig:tril} Scalar sum of jets and leptons transverse momenta in the 3$\ell$ channel's signal region. Background shown after the fit to the data under the background-only hypothesis. The last bin contains overflow.}
\end{figure}

 
In all cases there will be a need to select events compatible with a VLQ decay through a $Z$ boson and a third generation quark, so all channels will select at least two leptons with opposite-sign and same flavor with an invariant mass within 10~GeV of the $Z$ boson mass. There will be a requirement of at least 1 \bjet so that decays to a $b$ or a $t$ quark are selected. In the dilepton channels the large-R jet multiplicity splitting follows. In both cases a cut in the $Z$ candidate $\pt$ at 250~GeV is applied to suppress background contributions. A cut in the scalar sum of jets $\pt$ ($\HT$) is also applied, with 800~GeV being the threshold in the \dilres~channel and 1150~GeV in the \dilboos~channel. In \Fig{\ref{fig:dilres}} the $\HT$ distributions after the aforementioned selection is applied are shown. Two signal regions are used in the fit, splitting events into 0 or 1 large-R jets, taking advantage of the different signal-to-background ratios in each case. $Z$+jets and $\ttbar$ are the main backgrounds and a control region for each of the backgrounds was designed.\par 
Figure \ref{fig:dilboos} shows the discriminant variable, $m_{Zb}$, of the \dilboos~channel. It consists of the invariant mass of the $Z$ boson candidate and the highest $\pt$ \bjet. This is a reconstruction of the vector-like $B$. In the vector-like $T$ case it is not as good of a reconstruction due to the missing $W$ boson but it is a reasonable compromise. Both dilepton channels have the same main backgrounds and the \dilboos~channel also has two corresponding control regions.\par 
In the \trilep~channel after the requirement of a $Z$ boson candidate and at least 1 \bjet a cut on the $Z$ candidate $\pt$ is applied at 200~GeV. The discriminant variable is the $\ST$ of jets and leptons and it can be seen in Figure \ref{fig:tril}. The main backgrounds for this channel are dibosons and $\ttbar$+X, both of which have its own control region.\par 
All of the aforementioned thresholds were obtained by optimizing the expected VLQ lower mass limits. All the figures are shown after the fits of background prediction to data data in a background only hypothesis.
%%%%%%%%%%%%%%%%%%%%%%%%%%%%%%%%%%%%%%%%%%%%%%%%%%%%%%%%%%%%%%%%%%%%%%%%%
%%
%%   use this format to include an .eps figure into your paper
%%
%\begin{figure}[htb]
%\centering
%\includegraphics[height=1.5in]{magnet}
%\caption{Plan of the magnet used in the mesmeric studies.}
%\label{fig:magnet}
%\end{figure}
%%%%%%%%%%%%%%%%%%%%%%%%%%%%%%%%%%%%%%%%%%%%%%%%%%%%%%%%%%%%%%%%%%%%%%%%%%%
\section{Results}
\begin{figure}[h]
\centering
\includegraphics[height=1.2in]{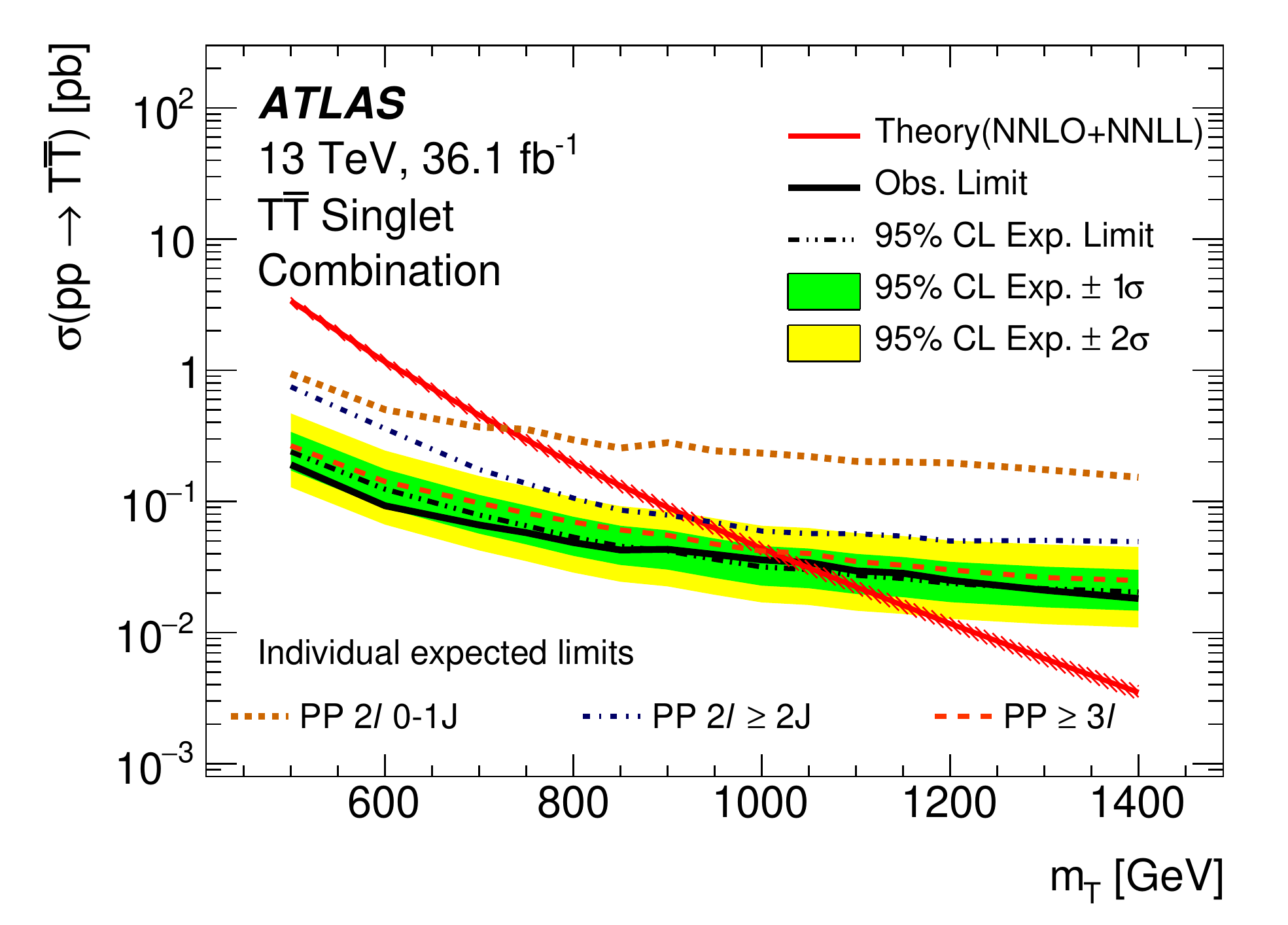}
\includegraphics[height=1.2in]{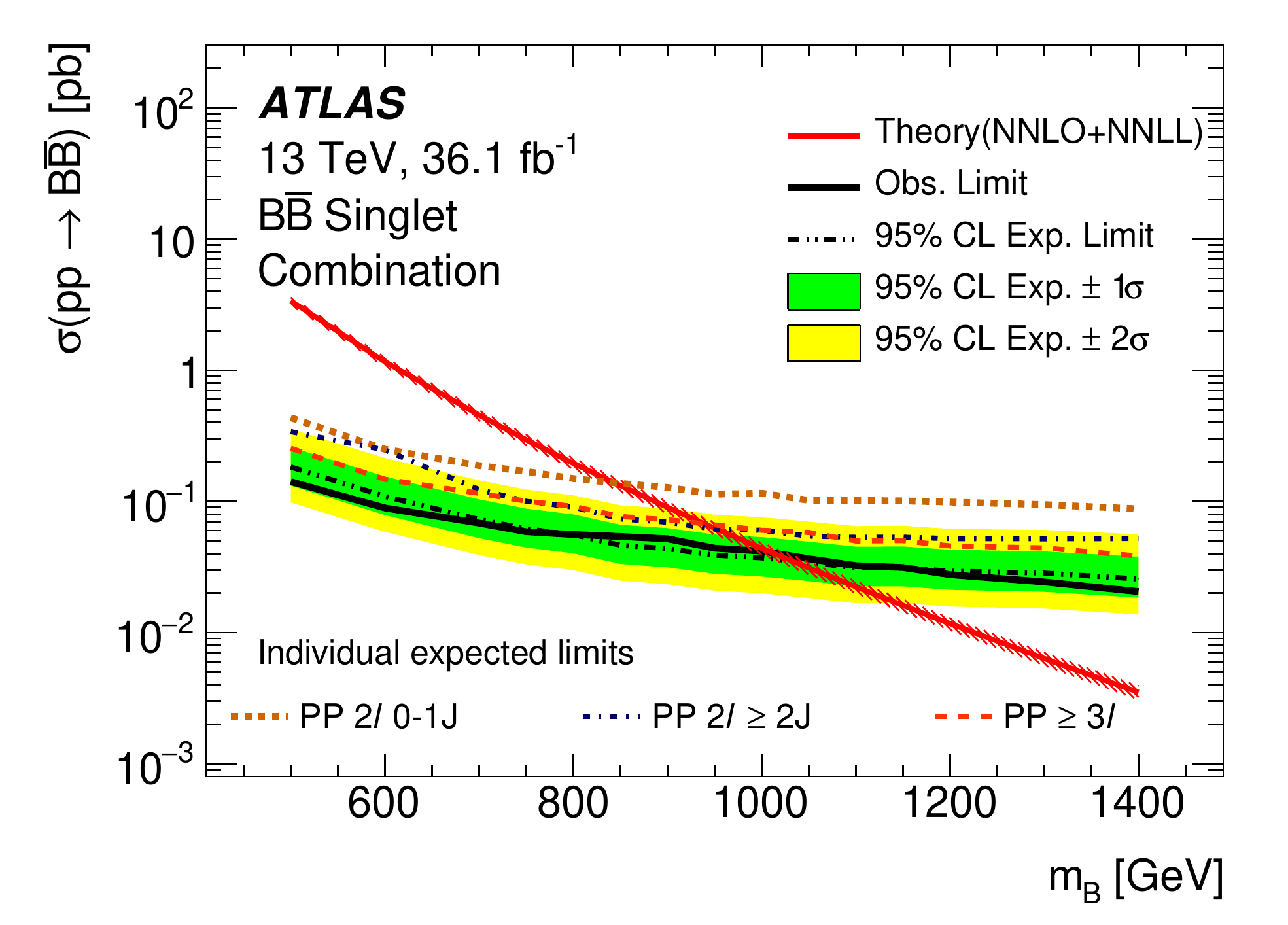}\\
\includegraphics[height=1.2in]{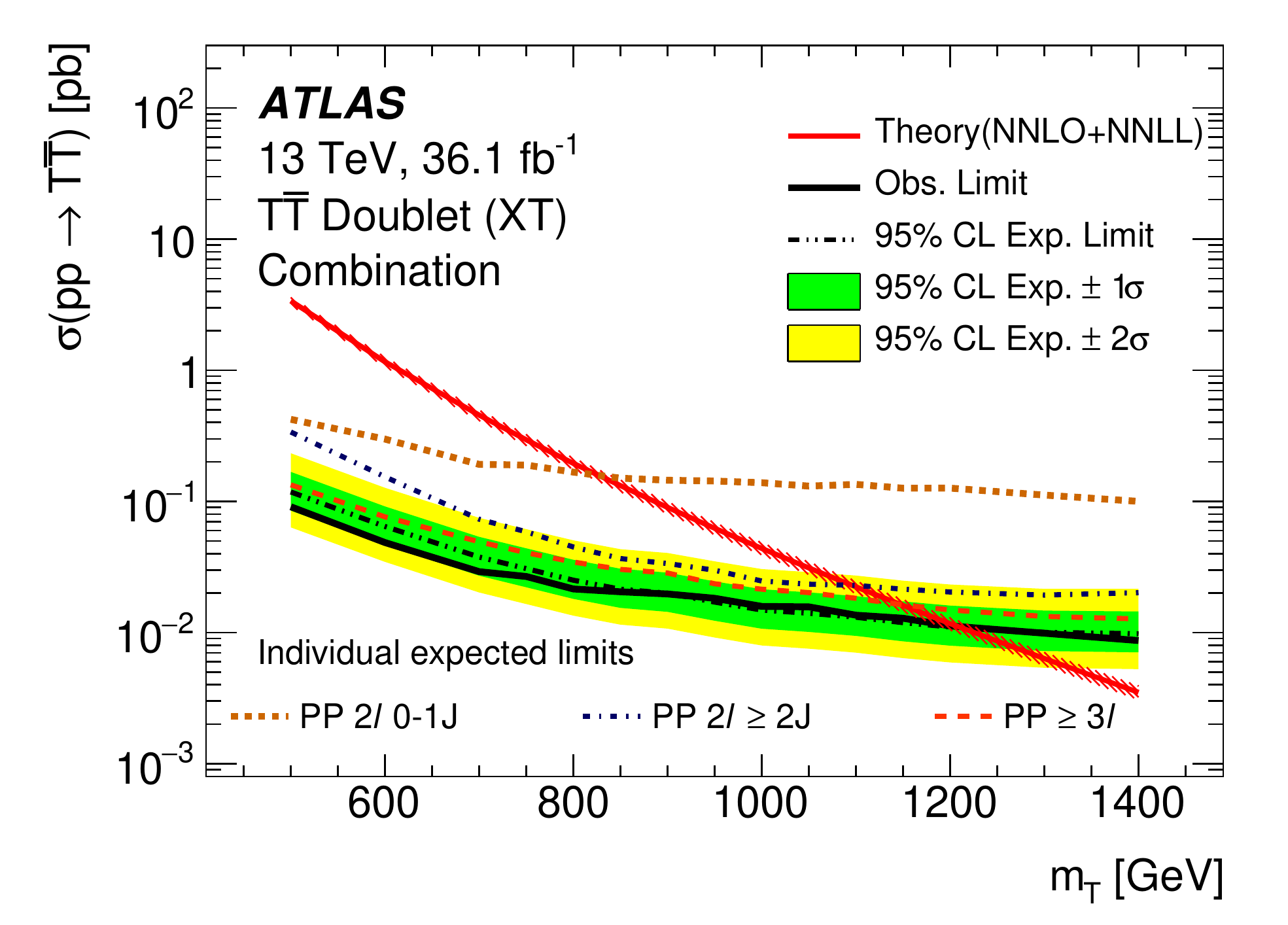}
\includegraphics[height=1.2in]{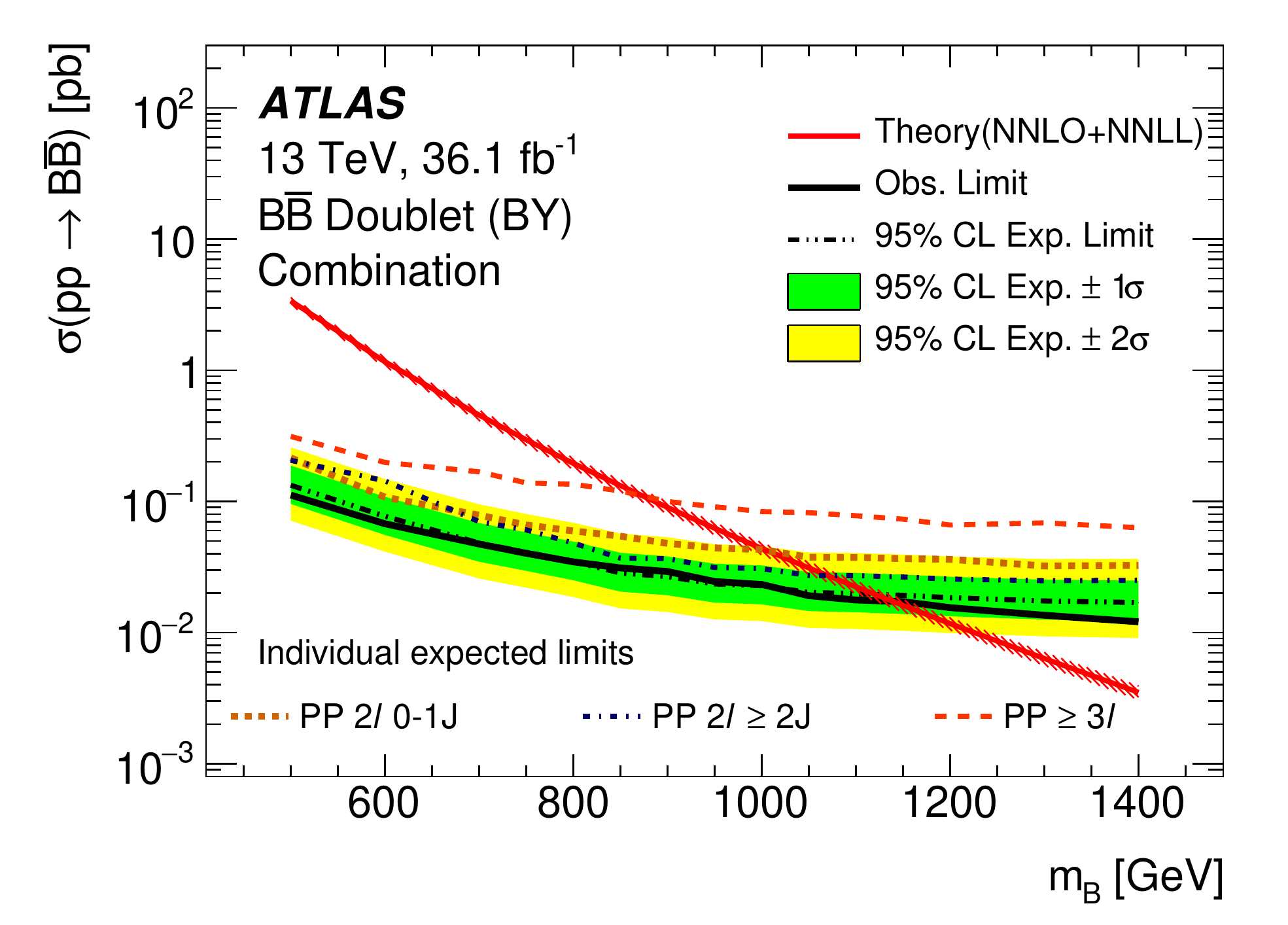}
\caption{Upper limits at 95$\%$ CL on the vector-like quark pair-production cross-section. Expected limits shown for each individual channel and for the combination of all channels. Observed limits are shown for the combination. The expected cross-sections, with their uncertainties, are shown in the red lines.}
\label{fig:lims}
\end{figure}

The three optimized channels were used in a statistical combination in order to increasing the overall sensitivity. A profile likelihood fit using the CL$_s$ method was performed to obtain 95$\%$ CL lower mass limits. The singlet and doublet hypothesis were tested for the vector-like $B$ and $T$. These results can be seen in Figure \ref{fig:lims}. For the singlet hypothesis the observed (expected) limits are 1030 (1060)~GeV for $T$ and 1010 (1030)~GeV for $B$. For the doublet hypothesis the observed (expected) limits are 1210 (1210)~GeV for $T$ and 1140 (1120)~GeV for $B$. In Figure \ref{fig:lims} the observed upper limits for VLQ pair-production cross-section are shown. The individual channel lines show their expected limits. It can be seen that the combination significantly improves the results with respect to the best channel in each scenario.\par
\begin{figure}
\centering
\begin{subfigure}{0.7\textwidth}
\includegraphics[width=0.491\textwidth]{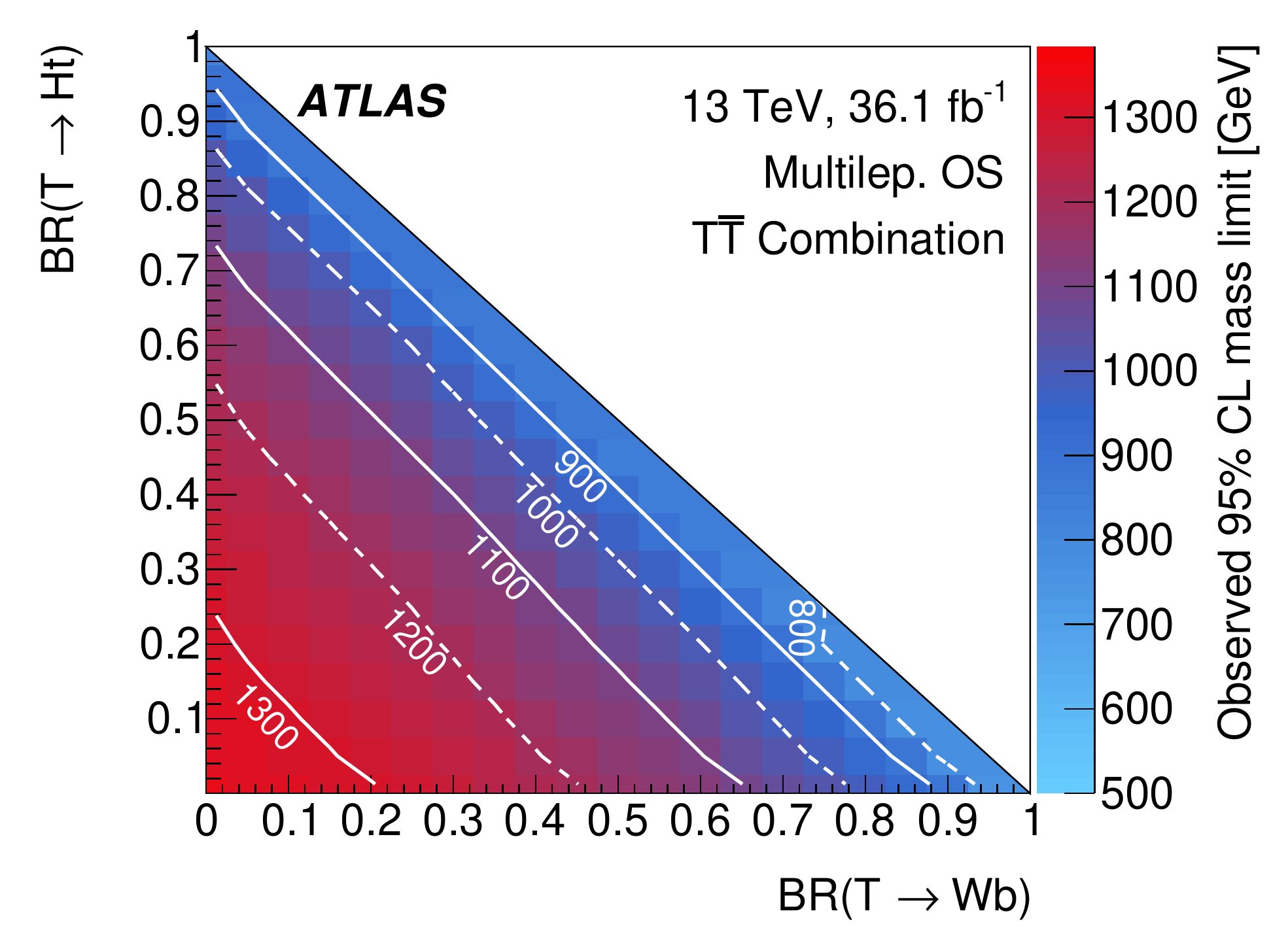}
\includegraphics[width=0.491\textwidth]{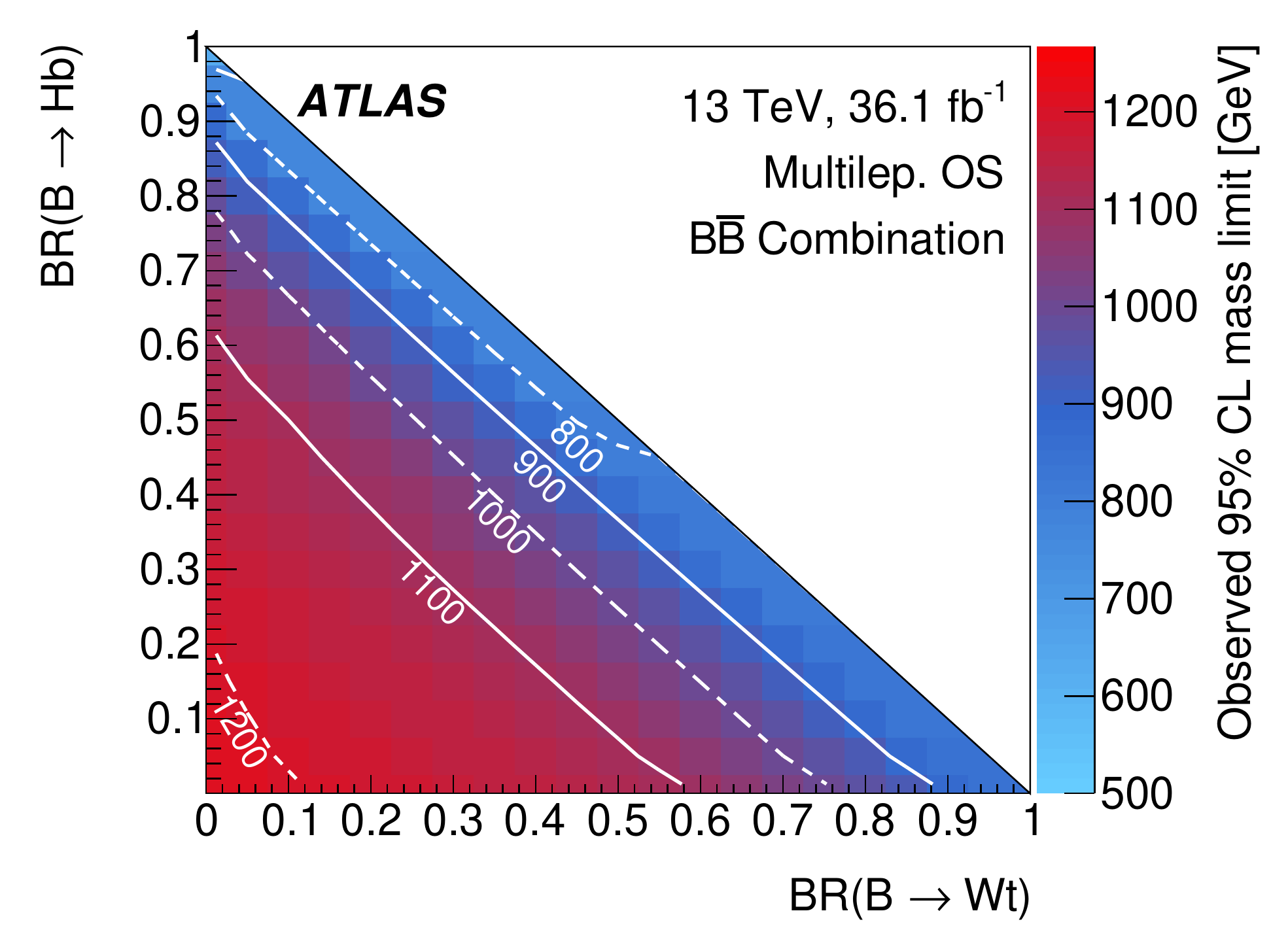}
\vspace*{-0.08\textwidth}
\caption{}
\label{fig:br_scan}
\end{subfigure}\\
\vspace*{0.03\textwidth}
\begin{subfigure}{0.7\textwidth}
\includegraphics[width=0.491\textwidth]{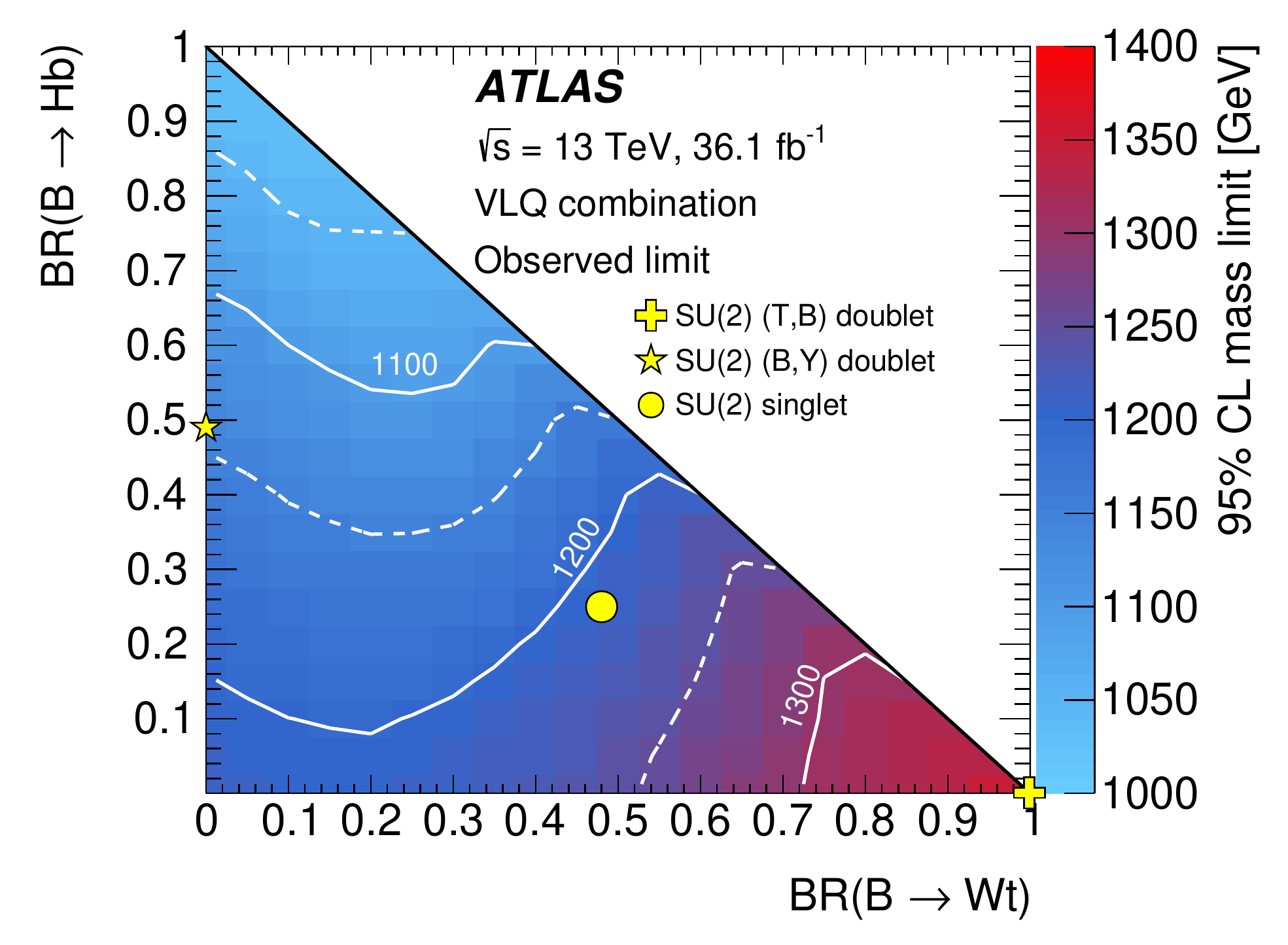}
\includegraphics[width=0.491\textwidth]{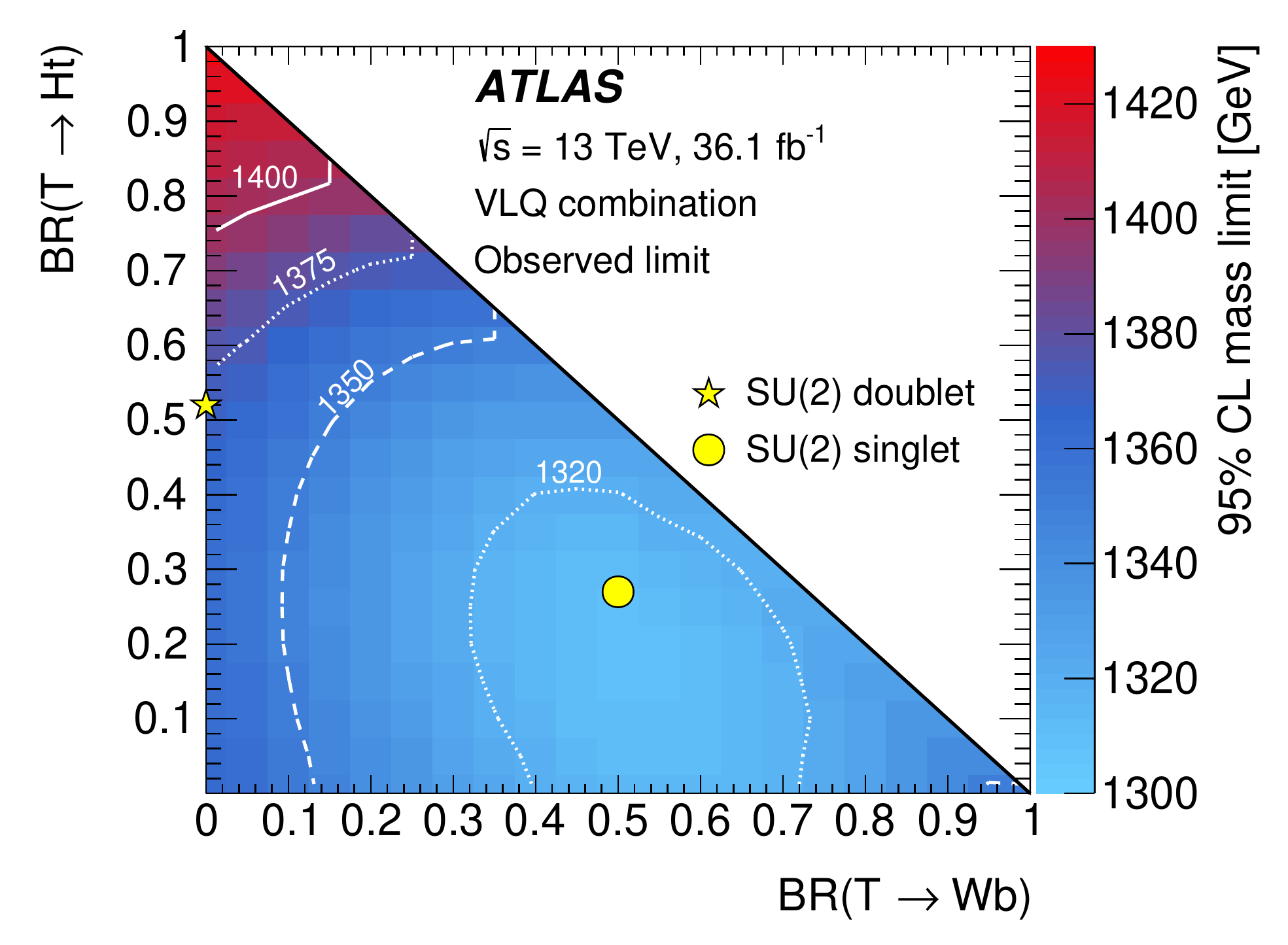}
\vspace*{-0.08\textwidth}
\caption{}
\label{fig:br_scan_combo}
\end{subfigure}
\caption{Observed 95$\%$ CL lower limits from the pair-production combination on the mass of vector-like quarks for all branching ratio combinations, adding up to unity for: \subref{fig:br_scan} the analysis in presented in this document; \subref{fig:br_scan_combo} combination of all pair-production analysis. Fixed mass values are shown as contours with white lines.}
\end{figure}

 
The results for each branching ratio possibility are shown in Figure \ref{fig:br_scan}, assuming that the branching ratios to $Z$, $W$ and $H$ add up to unity. It can be seen that the sensitivity increases in the $Z$ corner, as expected.\par 
The analysis described in this proceedings is integrated in a set of pair-production searches. Each targets a different final state with the goal of covering most of the BR scenarios. The specific final states will present different challenges so each search is individually optimized and they are all statistically combined afterwards, resulting in what is shown in Figure \ref{fig:br_scan_combo}\cite{combo}. When comparing the combination to the analysis presented in this poster there is an improvement of 270 (160)~GeV for the singlet (doublet) $T$ and 200 (10) for the singlet (doublet) $B$.
%\begin{figure}[h]
%\centering
%\includegraphics[height=1.5in]{figures/figaux_08.png}
%\caption{Observed (filled) and expected (dashed) 95$\%$ CL exclusion in the branching ratio plane for the different pair-production analyses.}
%\label{fig:comb}
%\end{figure}
 
\end{document}

%% file: econfmacros.tex
\def\beq{\begin{equation}}
\def\eeq#1{\label{#1}\end{equation}}
\def\eeqn{\end{equation}}

\def\beqa{\begin{eqnarray}}
\def\eeqa#1{\label{#1}\end{eqnarray}}
\def\eeqan{\end{eqnarray}}

\let\bar=\overbar

\def\Dslash{\not{\hbox{\kern-4pt $D$}}}
\def\dslash{\not{\hbox{\kern-2pt $\del$}}}

\def\msb{{\bar{\ssstyle M \kern -1pt S}}}